\documentclass[10pt,twocolumn]{article} 
\usepackage{amsmath,amsfonts}
\usepackage{algorithmic}
\usepackage{algorithm}
\usepackage{array}
\usepackage[caption=false,font=normalsize,labelfont=sf,textfont=sf]{subfig}
\usepackage{textcomp}
\usepackage{stfloats}
\usepackage{url}
\usepackage{verbatim}
\usepackage{graphicx}
\usepackage{cite}
\usepackage[colorlinks=true,linkcolor=black,anchorcolor=black,citecolor=black,filecolor=black,menucolor=black,runcolor=black,urlcolor=black]{hyperref}

\usepackage[top=2.4cm,bottom=2.4cm,left=2.4cm,right=2.4cm]{geometry}

\begin{document} 

\title{A Noise Model for Multilayer Graded-Bandgap Avalanche Photodiodes}

\author{Ankitha E. Bangera \\
\\
\it{Department of Electrical Engineering,} \\
\it{Indian Institute of Technology Bombay,} \\
\it{Mumbai$-$400076, India} \\
E-mail(s): ankitha\_bangera@iitb.ac.in; ankitha.bangera@iitb.ac.in}
\date{}

\maketitle
\thispagestyle{empty}

\begin{abstract}
Multilayer graded-bandgap avalanche photodiodes (APDs) are the future deterministic photomultipliers, owing to their deterministic amplification with twofold stepwise gain via impact ionization when operated in the staircase regime. Yet, the stepwise impact ionization irregularities worsen as the number of steps increases. These irregularities in impact ionization are the major source of noise in these APDs. These solid-state devices could replace conventional silicon photomultiplier tubes if they are carefully studied and designed. A noise model for multistep staircase APDs, considering equal stepwise ionization probabilities is previously reported. However, we derive a generalized noise model for multilayer graded-bandgap APDs, applicable for all operating biases, which include the sub-threshold, staircase, and tunnelling breakdown regimes. Moreover, the previous noise model's expression for the total excess noise factor in terms of ionization probabilities of the multistep staircase APD follows Friis's total noise factor. However, we demonstrate that our derived expression matches Bangera's correction to Friis's total noise factor.
\end{abstract}

\section{Introduction}
\par{Photomultiplier tubes (PMTs) \cite{ref1} are most often used for deterministic photomultiplication, owing to their extremely high gain and low noise \cite{ref2}. PMTs consist of a photocathode for the emission of photoelectrons (primary electrons), a focusing grid, an array of metallic dynodes which enable amplification by emitting Poisson-distributed secondary electrons, and an anode for collection of electrons at the output, all enclosed in a vacuum glass tube \cite{ref1}. These PMTs have several applications in photon counting \cite{ref1,ref3,ref4}, spectroscopy \cite{ref1,ref5,ref6}, high energy physics \cite{ref1,ref7,ref8}, radiation measurement \cite{ref1}, electron microscopy \cite{ref1,ref9}, and so on. Some biomedical applications of PMTs include positron emission tomography, in-vitro assay, computed radiography, and gamma scintigraphy \cite{ref1,ref10,ref11}. However, these devices are large in size (approximately a few cm), fragile, expensive, and operated at high voltages of greater than 1kV in a vacuum \cite{ref2}.}

Although the conventional avalanche photodiodes (APDs) are alternate solid-state devices for PMT applications such as photon counting \cite{ref12,ref13,ref14,ref15,ref16,ref17}, spectroscopy \cite{ref18,ref19,ref20}, biomedical \cite{ref21,ref22}, and so on; they produce a nonlinear output with high noise \cite{ref23,ref24,ref25,ref26}. Further attempts were made to reduce the excess noise and improve the gain in conventional APDs by fabricating quantum dot resonant tunelling diode based APDs \cite{ref13,ref27}, waveguide-integrated Germanium APDs \cite{ref28}, nanowire APDs \cite{ref29,ref30}, separate absorption-multiplication (SAM) APDs using tunable direct bandgap digital alloys such as AlInAsSb \cite{ref31,ref32,ref33}, and so forth. 

Recently, multilayer graded-bandgap APDs operated in their staircase operating regime have been the solid-state analogue of PMTs with deterministic amplification \cite{ref24,ref26,ref34,ref35,ref36,ref37,ref38,ref39}. The conduction band profile in the energy-position band-diagram of these APDs appears similar to a series of steps when the applied bias ranges within the staircase regime \cite{ref35,ref37,ref38,ref39,ref40,ref41}. Here, the function of the steps is similar to the metallic dynodes in a PMT. Thus, these APDs operated in the staircase regime may be referred to as multistep staircase APDs. Moreover, these solid-state devices have the advantages of micro-size, are low-cost, and are operated at low voltages. However, the irregularities in the stepwise impact ionization worsen as the number of steps increases \cite{ref24,ref26,ref39,ref41,ref42,ref43}. These solid-state devices could be a replacement for conventional silicon PMTs if they are carefully designed.  

This article first briefly discusses the existing theory and noise model of the conventional p-i-n APDs \cite{ref24,ref26,ref44}. We then propose a new noise model for a multilayer graded-bandgap APD. Owing to the irregularities in the stepwise impact ionization, we have derived a generalized noise model for a non-ideal $n$-layer graded-bandgap APD. The previous noise models \cite{ref35,ref41,ref43,ref45} for an $n$-step staircase APD provide the expression for the excess noise factor in terms of the number of steps when the stepwise impact ionizations are equal. However, our generalized model is applicable for all operating biases, which include the sub-threshold, staircase operating, and tunnelling breakdown regimes. Further, we present our model's simplified noise expression for $n$-step staircase APDs.  For validation, we compare the noise power ratios (NPRs) and the excess noise factors determined using our model with the previous models \cite{ref35,ref41,ref43,ref45}, when the stepwise ionization probabilities are equal. Then we discuss the inter-dependent variations of ionization probabilities, NPRs, noise current ratios (NCRs), excess noise factors, and staircase gains of 1-step, 2-step, and 3-step staircase APDs.

\section{Theory and Noise Model}

\subsection{Conventional p-i-n APD}

The theory and noise model of a conventional p-i-n APD is well known in the literature \cite{ref24,ref26,ref44}. Let the input photocurrent of a conventional p-i-n APD be represented as,

\begin{equation}
\label{eqn_1}
i_{\text{ph}} = \sum_{\alpha=1}^{N_0}h(t-t_{\alpha})
\end{equation} 

Where $\alpha$ is the injected photoelectron count; $N_0$ is the total number of charges generated by the incident radiation or the total injected photoelectrons; $h(t-t_{\alpha})$ is the pulse function that defines the input photocurrent corresponding to the injected photoelectron generated by the pulse of radiation incident at time $t_{\alpha}$, such that, $\int h(t) dt=q$, where $q$ is the electron charge. 

Then, the time-dependent output current of the conventional p-i-n APD, neglecting the dark current will be, 

\begin{equation}
\label{eqn_2}
i(t)=\sum_{\alpha=1}^{N_0}M_{\text{C}}h(t-t_{\alpha}-t_{\text{o}})
\end{equation} 

Where $t_{\text{o}}$ is the time taken for the photoelectron to reach the output, and $M_{\text{C}}$ is the avalanche gain of the conventional p-i-n APD for each pulse of charge. Here, $M_{\text{C}}=1$ when the device is operated in the unity gain mode regime such that the incident photon count is equal to the number of photoelectrons generated and collected at the output, which implies that no carrier multiplication happens \cite{ref26}. 

The noise current spectral intensity (A$^2$Hz$^{-1}$) of this conventional APD \cite{ref26,ref44}, neglecting the dark current is given by, 

\begin{equation}
\label{eqn_3}
\begin{aligned}
S_\text{C}^I (f) &= 2q \langle M_\text{C}^2\rangle I_0 \\
&= 2q \langle M_\text{C}\rangle ^2 F(M_\text{C}) I_0
\end{aligned}
\end{equation}

Where $q$ is the electron charge which is a constant, $F(M_{\text{C}}) \equiv \frac{\langle M_\text{C}^2\rangle}{\langle M_\text{C}\rangle ^2}$ is the excess noise factor of the conventional p-i-n APD, $I_0= \langle N_0 \rangle |H(f)|$ is the unity gain photocurrent. 

The spectral noise current (A$\sqrt{\textrm{Hz}}$) is defined as the square-root of the noise current spectral intensity \cite{ref26,ref44}. Therefore, the spectral noise current of this conventional APD is given by, 

\begin{equation}
\label{eqn_4}
\begin{aligned}
\sigma_\text{C}^I (f) &= \sqrt{2q \langle M_\text{C}^2\rangle I_0} \\
&=\sqrt{2q \langle M_\text{C}\rangle ^2 F(M_\text{C}) I_0}
\end{aligned}
\end{equation}

The noise power spectral density (W$\cdot$Hz$^{-1}$) is defined as the product of the noise current spectral intensity and the noise figure analyser's AC load resistance ($R_\text{L}$) \cite{ref26}. For this conventional APD, the noise power spectral density is given by, 

\begin{equation}
\label{eqn_5}
\begin{aligned}
S_\text{C}^P (f) &= 2q \langle M_\text{C}^2\rangle I_0 R_\text{L} \\
&=2q \langle M_\text{C}\rangle ^2 F(M_\text{C}) I_0 R_\text{L}
\end{aligned}
\end{equation}

\subsection{A generalized noise model for an \textit{n}-layer graded-bandgap APD}

\begin{figure}[!t]
\centering
\includegraphics[width=3in]{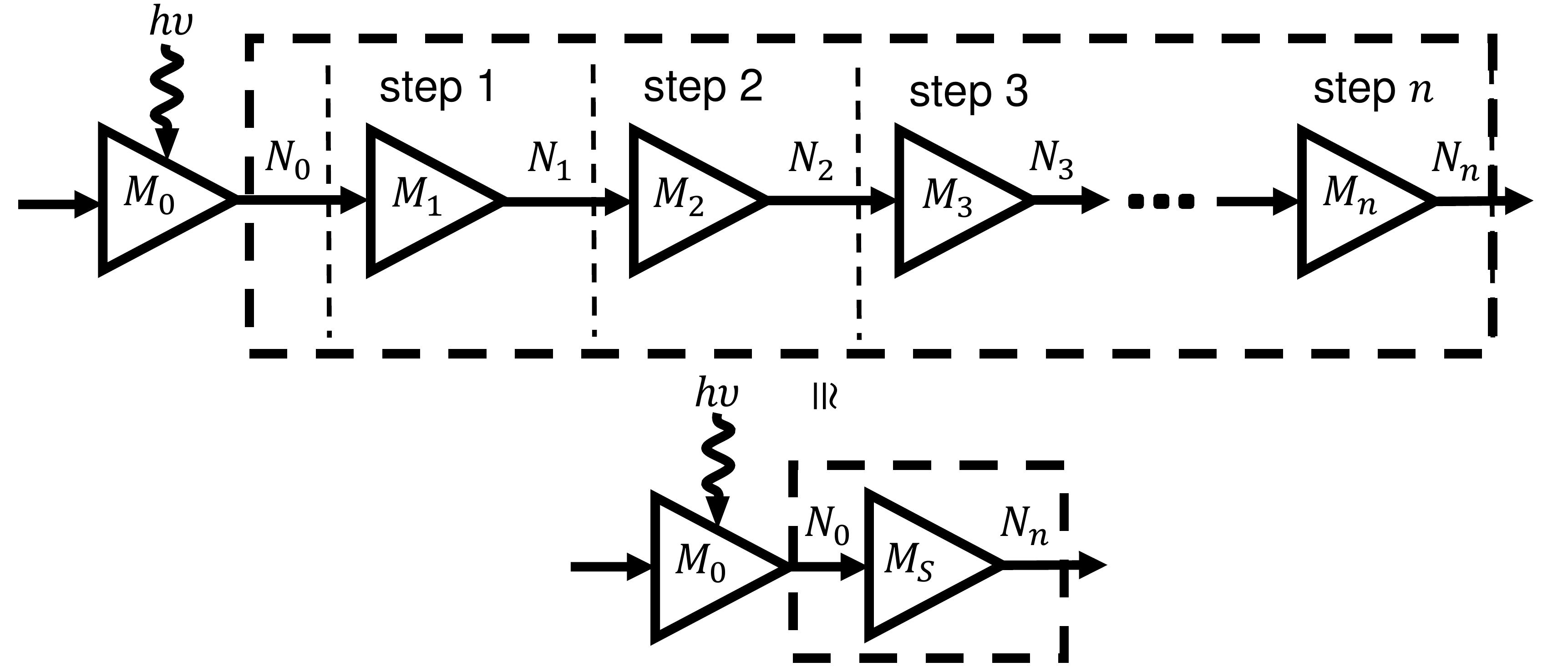}
\caption{Block diagram of an $n$-layer graded-bandgap APD.}
\label{fig_1}
\end{figure}

The block diagram shown in Fig.~\ref{fig_1} elaborates the amplification of the photo-generated electrons (photoelectrons) in an $n$-layer graded-bandgap APD. Here, the first amplifier with gain $M_0$ corresponds to the photoelectrons, and $M_0=1$ indicates that each incident photon generates only one photoelectron and no carrier multiplication happens in the absorption region of the APD. Moreover, the band structure of an $n$-layer graded-bandgap APD appears discontinuous due to the heterojunction interfaces, forming step-like structures when biased, especially in the staircase regime \cite{ref35,ref38}. This article refers to these junction discontinuities as steps, even when the devices are unbiased. Therefore, the junction/step gains are represented as $M_x$ corresponding to the gains at junction/step `$x$'. From the block diagram, it is clear that the total step gain (total multiplication gain) of the $n$-layer graded-bandgap APD can be defined as the fraction of the total number of electrons at the output of step `$n$' ($N_n$) and the total number of input photoelectrons ($N_0$). This is also equal to the product of all the step gains. Therefore, the total step gain is represented as,

\begin{equation}
\label{eqn_6}
M_\text{S}=\frac{N_n}{N_0}=M_1M_2M_3...M_n=\prod_{x=1}^{n}M_x
\end{equation}

If $h(t-t_{\alpha})$ is a pulse-shaped function that represents input photocurrent corresponding to the injected photoelectron generated by the pulse of radiation incident at time $t=t_{\alpha}$, then, $h(t-t_{\alpha}-t_x)$ is the pulse-shaped function shifted by a time instant `$t_x$'. In comparison to the conventional p-i-n APD, the time-dependent output current of an $n$-layer graded-bandgap APD, neglecting the dark current is given by,

\begin{equation}
\label{eqn_7}
\begin{aligned}
i(t) &= \sum_{\alpha=1}^{N_0} M_{\text{S}} M_0 \delta(t-t_n)*h(t-t_{\alpha}) \\
&= \sum_{\alpha=1}^{N_0} M_{\text{S}} M_0 h(t-t_{\alpha}-t_n) \\
&= \sum_{\alpha=1}^{N_0} \left[ \prod_{x=1}^{n}M_x \right] M_0 h(t-t_{\alpha}-t_n)
\end{aligned}
\end{equation} 

Where `$n$' is the total number of layers in the $n$-layer graded-bandgap APD, $\delta(t-t_n)*h(t-t_{\alpha})$ represents the time-shifted pulse-shaped function and `$*$' indicates convolution operator. 

The noise current spectral intensity of this $n$-layer graded-bandgap APD is

\begin{equation}
\label{eqn_8}
\begin{aligned}
S_\text{S}^I (f) &=2q \langle M_0^2\rangle \langle M_\text{S}^2\rangle I_0 \\
&=2q \langle M_0\rangle ^2 F(M_0) \langle M_\text{S}\rangle ^2 F(M_\text{S}) I_0
\end{aligned}
\end{equation}

Where $F(M_0)$ is the excess noise factor corresponding to the gain $M_0$ and $F(M_\text{S})$ is the excess noise factor corresponding to the gain $M_\text{S}$ of the $n$-layer graded-bandgap APD. 

The spectral noise current of this $n$-layer graded-bandgap APD is given by, 

\begin{equation}
\label{eqn_9}
\begin{aligned}
\sigma_\text{S}^I (f) &=\sqrt{2q \langle M_0^2\rangle \langle M_\text{S}^2\rangle I_0} \\
&=\sqrt{2q \langle M_0\rangle ^2 F(M_0) \langle M_\text{S}\rangle ^2 F(M_\text{S}) I_0}
\end{aligned}
\end{equation}

The noise power spectral density of this $n$-layer graded-bandgap APD is,

\begin{equation}
\label{eqn_10}
\begin{aligned}
S_\text{S}^P (f) &=2q \langle M_0^2\rangle \langle M_\text{S}^2\rangle I_0 R_{\text{L}} \\
&=2q \langle M_0\rangle ^2 F(M_0) \langle M_\text{S}\rangle ^2 F(M_\text{S}) I_0 R_{\text{L}}
\end{aligned}
\end{equation}

Further, this article defines the noise power ratio of the $n$-layer graded-bandgap APD ($\text{NPR}_{\text{S}_n}$) as the ratio of the noise power spectral density of the $n$-layer graded-bandgap APD to the noise power spectral density of the conventional p-i-n APD. When, $M_0=M_\text{C}$, we get, $F(M_0)=F(M_\text{C})$. Thus,

\begin{equation}
\label{eqn_11}
\begin{aligned}
\text{NPR}_{\text{S}_n}&=\frac{S_\text{S}^P (f)}{S_\text{C}^P (f)}\\
&=\frac{\langle M_0\rangle ^2 F(M_0) \langle M_\text{S}\rangle ^2 F(M_\text{S})}{\langle M_\text{C}\rangle ^2 F(M_\text{C})}\\
&=\langle M_\text{S}\rangle ^2 F(M_\text{S}) \\
&=\langle M_\text{S}^2\rangle 
\end{aligned}
\end{equation}

Similarly, we define the noise current ratio of the $n$-layer graded-bandgap APD ($\text{NCR}_{\text{S}_n}$) as the ratio of the spectral noise current of the $n$-layer graded-bandgap APD to the spectral noise current of the conventional p-i-n APD. This may also be defined as the square-root of the noise power ratio. Therefore,

\begin{equation}
\label{eqn_12}
\begin{aligned}
\text{NCR}_{\text{S}_n}&=\frac{\sigma_\text{S}^I (f)}{\sigma_\text{C}^I (f)}\\
&=\sqrt{\frac{\langle M_0\rangle ^2 F(M_0) \langle M_\text{S}\rangle ^2 F(M_\text{S})}{\langle M_\text{C}\rangle ^2 F(M_\text{C})}}\\
&=\sqrt{\langle M_\text{S}\rangle ^2 F(M_\text{S})}\\
&=\sqrt{\langle M_\text{S}^2\rangle}\\
&=\sqrt{\text{NPR}_{\text{S}_n}}
\end{aligned}
\end{equation}

In the new noise model proposed by us, let $X_x$ be a random variable for multiplication at step `$x$' that defines the number of extra electrons generated by a single electron at the input of each step, such that 

\begin{equation}
\label{eqn_13}
\begin{aligned}
X_x \sim 
\begin{cases}
0 & 1-p_{x1}-p_{x2}-...-p_{xm} \\
& [i.e., \text{No Ionization at step `}x\text{'} \\
&~~~~~~~~~~~~~~~~~~~~~~~~~~~~~~\implies X_x=0]\\
1 & p_{x1} \\ 
& [i.e., \text{Ionization at step `}x\text{'} \ni \text{it generates} \\ 
&~~~~~~~~~~~~~~~~~~~~~~~~~~~~~~~~~~~~~~~\text{1 e}^-]\\
2 & p_{x2} \\
& [i.e., \text{Ionization at step `}x\text{'} \ni \text{it generates} \\ 
&~~~~~~~~~~~~~~~~~~~~~~~~~~~~~~~~~~~~~~~\text{2 e$^-$s}]\\
\vdots & \vdots \\
m & p_{xm} \\
& [i.e., \text{Ionization at step `}x\text{'} \ni \text{it generates} \\ 
&~~~~~~~~~~~~~~~~~~~~~~~~~~~~~~~~~~~~~~~\text{`}m\text{' e$^-$s}]
\end{cases}
\end{aligned}
\end{equation}

Then, the generalized form of the time-dependent current at the output of the step `$n$', in terms of the random variable for multiplication `$X_x$' of an $n$-layer graded-bandgap APD, can be written as, 

\begin{equation}
\label{eqn_14}
\begin{aligned}
i(t)&=\sum_{\alpha=1}^{N_0}\left[\prod_{x=1}^{n}(1+X_x)\right] M_0h(t-t_{\alpha}-t_n)\\
&=\sum_{\alpha=1}^{N_0} \Biggl \{ \biggl[\prod_{x=1}^{n}(1+p_{x1}+2p_{x2}+...+np_{xm})\biggr] \\ 
&\hspace{3cm} \times M_0h(t-t_{\alpha}-t_n) \Biggr \}
\end{aligned}
\end{equation} 

Where the generalized equation for the total step gain may be represented as,

\begin{equation}
\label{eqn_15}
\begin{aligned}
M_{\text{S}} &=  \prod_{x=1}^{n}M_x = \prod_{x=1}^{n} (1+X_x) \\ 
&= \prod_{x=1}^{n}(1+p_{x1}+2p_{x2}+...+np_{xm}) 
\end{aligned}
\end{equation} 

Similarly, the generalized equation for the noise power ratio and noise current ratio of the $n$-layer graded-bandgap APD may, respectively, be formulated as 

\begin{equation}
\label{eqn_16}
\begin{aligned}
\text{NPR}_{\text{S}_n}^{\text{new}} &= \bigg\langle \prod_{x=1}^{n}M_x^2 \bigg\rangle = \bigg\langle \prod_{x=1}^{n} (1+X_x)^2 \bigg\rangle \\ 
&= \bigg\langle \prod_{x=1}^{n}(1+p_{x1}+2p_{x2}+...+np_{xm})^2 \bigg\rangle 
\end{aligned}
\end{equation} 

and 

\begin{equation}
\label{eqn_17}
\begin{aligned}
\text{NCR}_{\text{S}_n}^{\text{new}} &= \sqrt{\bigg\langle \prod_{x=1}^{n}M_x^2 \bigg\rangle} = \sqrt{\bigg\langle \prod_{x=1}^{n} (1+X_x)^2 \bigg\rangle} \\ 
&= \sqrt{\bigg\langle \prod_{x=1}^{n}(1+p_{x1}+2p_{x2}+...+np_{xm})^2 \bigg\rangle} 
\end{aligned}
\end{equation} 

Here, since the $X_x$ is considered a random variable ranging from 0 to $m$ with different probabilities, this is a generalized noise model for a practical $n$-layer graded-bandgap APD applicable for all operating biases, which includes the sub-threshold, staircase operating, and tunneling breakdown regimes.

\section{Results and Discussion}

\subsection{Staircase operation of an \textit{n}-layer graded-bandgap APD}

An $n$-layer graded-bandgap APD operated in its staircase operation regime is called a multistep staircase APD or an $n$-step staircase APD. Considering that each incident photon generates only one photoelectron and no carrier multiplication happens in the absorption region of the APD, the photo-generated electron gain is expected to be $M_0=1$. Thus, the excess noise factor corresponding to $M_0$ will be $F(M_0)=1$. Therefore, the time-dependent output current of the $n$-step staircase APD will be, 

\begin{equation}
\label{eqn_18}
\begin{aligned}
i(t)&=\sum_{\alpha=1}^{N_0}M_{\text{S}}h(t-t_{\alpha}-t_n)\\
&=\sum_{\alpha=1}^{N_0}\left[\prod_{x=1}^{n}M_x\right]h(t-t_{\alpha}-t_n)
\end{aligned}
\end{equation} 

The noise power spectral density of this $n$-step staircase APD is, 

\begin{equation}
\label{eqn_19}
\begin{aligned}
S_\text{S}^P (f) &= 2q \langle M_\text{S}\rangle ^2 F(M_\text{S}) I_0 R_{\text{L}}\\
&= 2q \langle M_\text{S}^2\rangle I_0 R_{\text{L}}
\end{aligned}
\end{equation}

\begin{figure}[!t]
\centering
\includegraphics[width=3in]{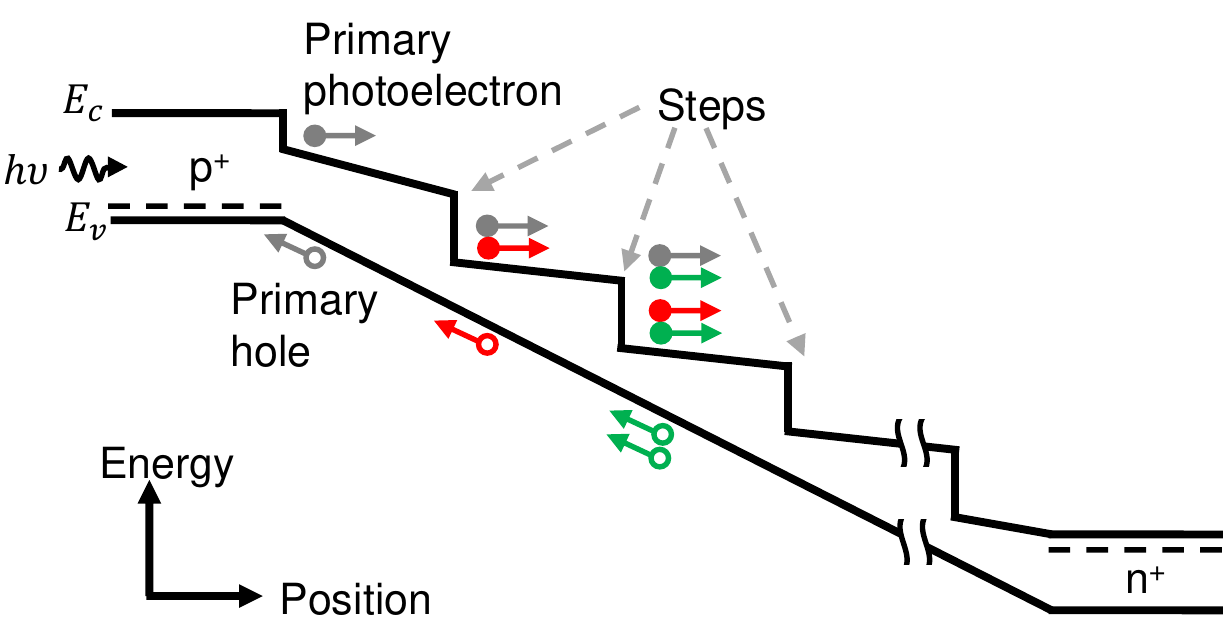}
\caption{Energy-position band-diagram of a multistep staircase APD.}
\label{fig_2}
\end{figure}

Fig.~\ref{fig_2} depicts the energy-position band-diagram of an ideal multistep staircase APD. Therefore, for an ideal $n$-step staircase APD, an electron at the input of each step will contribute to ionization by generating only one extra free electron. Thus, the gain at each step will be $M_x=(1+X_x)=2$, where `$x$' is the step number; and the total gain of the staircase APD will be a deterministic gain equal to  $M_\text{S}=(1+X_x)^n=2^n$. Here, the $X_x$ will always be equal to 1, \textit{i.e.}, the probability ($p_x$) of the event $X_x=1$ is `one.'

Therefore, the noise power spectral density of such an $n$-step staircase APD with equal stepwise ionization probability $p_x=1$ at all steps such that all input electrons at every step generates only one extra electron is, 

\begin{equation}
\label{eqn_20}
\begin{aligned}
S_\text{S}^P (f) &= 2q\bigg\langle \prod_{x=1}^{n}(1+X_x)^2 \bigg\rangle I_0 R_\text{L} = 2q (2)^{2n} I_0 R_\text{L} 
\end{aligned}
\end{equation}

Further, comparing the equations (\ref{eqn_19}) and (\ref{eqn_20}), the noise power ratio of the $n$-step staircase APD in terms of $X_x$ is formulated as,

\begin{equation}
\label{eqn_21}
\text{NPR}_{\text{S}_n}^\text{new}= \langle M_\text{S}\rangle ^2 F(M_\text{S}) = \langle M_\text{S}^2\rangle = \bigg\langle \prod_{x=1}^{n}(1+X_x)^2 \bigg\rangle
\end{equation}

However, in the case of a practical $n$-step staircase APD, an electron at the input of step `$x$' would generate one or more extra electrons when ionized, or the input electrons may not ionize \cite{ref35,ref38,ref39,ref41,ref43,ref45}. Furthermore, all the steps may not have equal ionization probabilities, owing to the non-identically manufactured APD heterojunctions, the device operating bias, and so on \cite{ref39}. Thus, $X_x$ indicated in equation (\ref{eqn_13}) would provide a generalized solution for all the irregularities included. Moreover, from the literature \cite{ref35,ref38,ref39,ref41,ref43,ref45}, it is known that the probability of input electrons generating more than two extra electrons at each step is quite low when the devices are operated in the staircase operating regime. Therefore, simplifying the equations and neglecting the ionization events with lower probabilities at step `$x$', we consider that an electron at the input of step `$x$' would generate only one extra electron when ionized and no extra electrons if not ionized. Then, $X_x$ could be considered as a random variable with just two possibilities 0 or 1 with probabilities $(1-p_x)$ or $p_x$, respectively. Therefore,

\begin{equation}
\label{eqn_22}
\begin{aligned}
X_x \sim 
\begin{cases}
0 & (1-p_{x}) \\
& [i.e., \text{No Ionization at step `}x\text{'} \implies X_x=0]\\
1 & p_{x} \\ 
& [i.e., \text{Ionization at step `}x\text{'} \ni \text{1 e$^-$ generated}]\\
\end{cases}
\end{aligned}
\end{equation}

The simplified time-dependent current of a staircase APD at the output of step `$n$' will be,

\begin{equation}
\label{eqn_23}
\begin{aligned}
i(t) &= \sum_{\alpha=1}^{N_0}\left[\prod_{x=1}^{n}(1+X_x)\right]\delta(t-t_n)*h(t-t_{\alpha}) \\
&= \sum_{\alpha=1}^{N_0}\left[\prod_{x=1}^{n}(1+p_x)\right] h(t-t_{\alpha}-t_n) 
\end{aligned}
\end{equation}

Moreover, the noise power ratio of the proposed new noise model is,

\begin{equation}
\label{eqn_24}
\begin{aligned}
\text{NPR}_{\text{S}_n}^\text{new} &= \bigg\langle \prod_{x=1}^{n}(1+X_x)^2 \bigg\rangle \\
&= \prod_{x=1}^{n}(1+p_x) \Biggl\{ 1+ \sum_{i=1}^{n}(2^2)^i \Biggl[ \sum_{j_1=1}^{[n-(i-1)]}\sum_{j_2=j_1+1}^{[n-(i-2)]} ...\\
& \hspace{1cm} \sum_{j_i=j_{i-1}+1}^{n}\frac{p_{j_1}}{(1-p_{j_1})}\frac{p_{j_2}}{(1-p_{j_2})} ... \frac{p_{j_i}}{(1-p_{j_i})} \Biggr]\Biggr\}
\end{aligned}
\end{equation} 

Equation (\ref{eqn_24}) may be simplified and rewritten as follows, 

\begin{equation}
\label{eqn_25}
\begin{aligned}
\text{NPR}_{\text{S}_n}^\text{new} &= 1+ \sum_{i=1}^{n}(3)^i \Biggl[ \sum_{j_1=1}^{[n-(i-1)]}\sum_{j_2=j_1+1}^{[n-(i-2)]} ...\\
&\hspace{1cm} \sum_{j_i=j_{i-1}+1}^{n}p_{j_1}p_{j_2} ... p_{j_i}\Biggr]
\end{aligned}
\end{equation} 

The derivation and proof of the above equations (\ref{eqn_24}) and (\ref{eqn_25}) are included in Appendix~\ref{Appx_1}. 

\subsection{A comparison with the previous noise models for staircase APDs}

From one of the previous noise models \cite{ref35,ref41,ref43}, it is reported that if `$\delta$' is the fraction of electrons at the input of each step that do not impact-ionize, then the total step gain of the $n$-step staircase APD is $M_\text{S}^\text{prev1}=(2-\delta)^n$ and its excess noise factor is

\begin{equation}
\label{eqn_26}
\begin{aligned}
F(M_\text{S},\delta)^\text{prev1} &= 1+\sum_{x=1}^{n}\Biggl( \frac{1}{(2-\delta)^x}.\frac{\delta(1-\delta)}{(2-\delta)} \Biggr)\\
&=1+\frac{\delta(1-(2-\delta)^{-n})}{(2-\delta)}
\end{aligned}
\end{equation} 

Another literature on the noise model \cite{ref39,ref45} reports that if `$p$' is the ionization probability at each step, then the excess noise factor $F(M_\text{S})^\text{prev2}$ is, 

\begin{equation}
\label{eqn_27}
\begin{aligned}
F(M_\text{S},p)^\text{prev2}&=1+\frac{(1-p)(1-(1+p)^{-n})}{(1+p)}\\
&=F(M_\text{S},\delta)^\text{prev1}=F(M_\text{S})^\text{prev}
\end{aligned}
\end{equation} 

Here, both the previous models are equivalent where the $F(M_\text{S},p)^\text{prev2}$ will be the same as $F(M_\text{S},\delta)^\text{prev1}$ when the `$\delta$' is replaced by $(1-p)$, and the total step gain of the $n$-step staircase APD in terms of `$p$' will be $M_\text{S}^\text{prev2}=(1+p)^n$. As reported in the literature \cite{ref35,ref39,ref41,ref43,ref45}, the formulas for the excess noise factor of $n$-step staircase APDs in both the previous noise models follow Friis's formulas for excess noise factor of $n$-stage cascade networks \cite{ref46,ref47}. Further, according to the previous noise models \cite{ref35,ref39,ref41,ref43,ref45}, the noise power ratio in terms of `$p$' will be, 

\begin{equation}
\label{eqn_28}
\begin{aligned}
\text{NPR}_{\text{S}_n}^\text{prev}=(1+p)^{2n}\Biggl[ 1+\frac{(1-p)(1-(1+p)^{-n})}{(1+p)} \Biggr]
\end{aligned}
\end{equation} 

For comparison with the previous models, if our proposed new noise model's ionization probabilities at all the `$n$' steps are equal to `$p$,' then $M_\text{S}=\langle M_\text{S} \rangle=(1+p)^n$ is in accordance with the literature and previous models \cite{ref35,ref39,ref41,ref43,ref45}. However, from equation (\ref{eqn_25}), the noise power ratio of the proposed new noise model will be as formulated in equation (\ref{eqn_29}), and its proof is provided in Appendix~\ref{Appx_1}. 

\begin{equation}
\label{eqn_29}
\begin{aligned}
\text{NPR}_{\text{S}_n}^\text{new} = 1+\sum_{i=1}^{n}~^nC_i(3p)^i=\sum_{i=0}^{n}~^nC_i(3p)^i
\end{aligned}
\end{equation}

Therefore, the new noise model's expression for the excess noise factor $F(M_\text{S})^\text{new}$ is,

\begin{equation}
\label{eqn_30}
F(M_\text{S})^\text{new}=\frac{\langle M_\text{S}^2 \rangle}{\langle M_\text{S} \rangle^2}=\frac{\sum_{i=0}^{n}~^nC_i(3p)^i}{(1+p)^{2n}}
\end{equation}

\begin{figure*}[!t]
\centering
\includegraphics[width=6.3in]{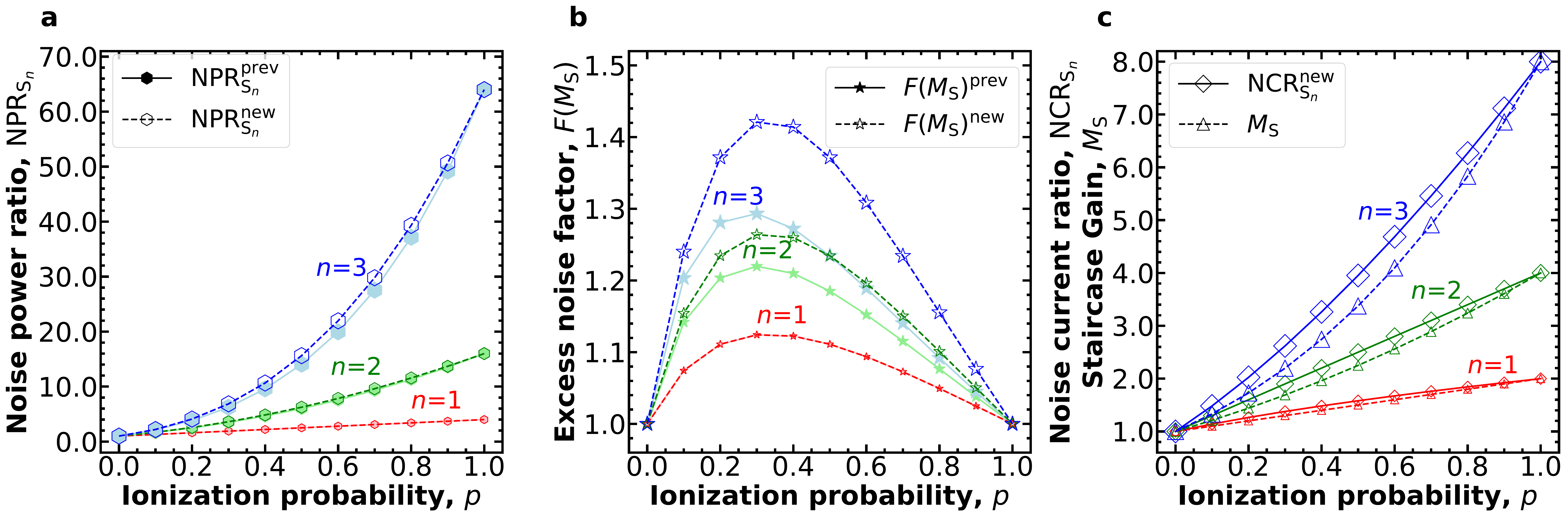}
\caption{A comparison of the (a) Noise power ratio and (b) Excess noise factor versus the ionization probability of the staircase APDs, with the total number of steps 1, 2, and 3 estimated using the previous noise models and the new noise model; and (c) The variation of noise current ratio and staircase gain with respect to the ionization probability of the staircase APDs, with the total number of steps 1, 2, and 3 estimated using the new noise model.}
\label{fig_3}
\end{figure*}

Fig.~\ref{fig_3}a compares the variation of the $\text{NPR}_{\text{S}_n}$ of the previous and the new noise models with the ionization probabilities. Further, Fig.~\ref{fig_3}a shows that for multistep staircase APDs with steps $n>1$ and ionization probabilities `$p$' in the range $0<p<1$, the $\text{NPR}_{\text{S}_n}^\text{prev} < \text{NPR}_{\text{S}_n}^\text{new}$. Since the expressions for the staircase gain (total step gain of a staircase APD) in the previous and new noise models remain the same, the difference in the $\text{NPR}_{\text{S}_n}$ is attributed to the difference in the total excess noise factors in the two models, as shown in Fig.~\ref{fig_3}b. Here, the excess noise factors $F(M_\text{S})$ of the multistep staircase APDs with steps $n=$ 1, 2, 3 can be considered as the total noise factors of cascade networks with the total number of stages $n=$ 1, 2, 3; with no externally added stage-wise noise. Thus, if the ionization probability is equal to one, then the stage-wise noise factors must be equal to one. However, suppose the ionization probabilities are less than one. In that case, there will be irregularities in the stage-wise multiplication factor, leading to the addition of internally generated stage-wise noise that is dependent on the ionization probabilities of that stage. These irregularities in ionization thus contribute to the stage-wise excess noise factors. Thus, in the case of staircase APDs with no externally added stage-wise noise, the stepwise noise factors ($F_x(M_\text{S})$) must be equal for all steps `$x$', with $F_x(M_\text{S})>$1 if $p\neq$1, 0.

From Fig.~\ref{fig_3}b, the previous model's excess noise factors $F(M_\text{S})^\text{prev}$ of the multistep staircase APDs with the total number of steps/stages `$n$' are in accordance with Friis's total noise factor formula for cascade networks \cite{ref46,ref47} given by,

\begin{equation}
\label{eqn_31}
F_{\text{T}_n}^{\text{Friis}} = F_1^{\text{Friis}}+\sum_{x=2}^{n}\left(\frac{F_x^{\text{Friis}}-1}{\prod_{y=1}^{(x-1)}M_y}\right) 
\end{equation}

An illustration of how the excess noise factors obtained using the previous noise models follow Friis's total noise factor formula for cascade networks is included in Appendix~\ref{Appx_2}. However, our new model's excess noise factors $F(M_\text{S})^\text{new}$ of the multistep staircase APDs agree with Bangera's total noise factor of $n$-stage cascade networks \cite{ref48} given by equation (\ref{eqn_32}) and is illustrated in Appendix~\ref{Appx_3}. 

\begin{equation}
\label{eqn_32}
F_{\text{T}_n}^{\text{Cor}} = \prod_{x=1}^{n}F_x^{\text{Cor}} 
\end{equation}

From the literature, it is known that Bangera's expression for the total noise factor of an $n$-stage cascade network in terms of the stage-wise noise factors is a correction to the corresponding Friis's formula \cite{ref48}. Thus, our new noise model for graded-bandgap $n$-step staircase APDs is an improved noise model. 

\begin{figure*}[!t]
\centering
\includegraphics[width=6.3in]{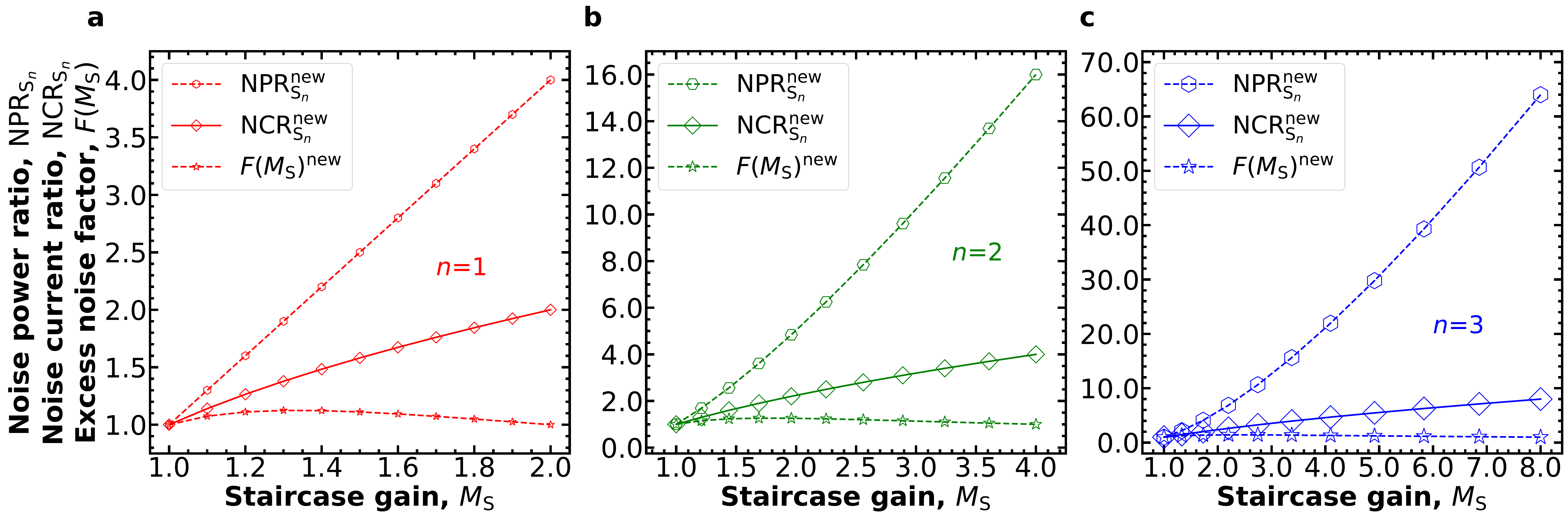}
\caption{The variation of the noise power ratio, noise current ratio, and excess noise factor versus the staircase gain of the $n$-step staircase APD, for steps (a) 1, (b) 2, and (c) 3 in accordance with our proposed noise model.}
\label{fig_4}
\end{figure*}

For better understanding and visualization of our proposed new noise model, the variation of the noise current ratio ($\text{NCR}_{\text{S}_n}$) and staircase gain ($M_\text{S}$) of the $n$-step staircase APD with the ionization probabilities is shown in Fig.~\ref{fig_3}c. We have also plotted the variation of the $\text{NPR}_{\text{S}_n}$, $\text{NCR}_{\text{S}_n}$, and $F(M_\text{S})$ versus $M_\text{S}$ of the $n$-step staircase APD, for steps 1, 2, and 3, shown in Fig.~\ref{fig_4}a,~\ref{fig_4}b, and~\ref{fig_4}c, respectively.

\section{Conclusion} 

We conclude that our generalized noise model for multilayer graded-bandgap APDs is applicable to estimate various noise expressions for all operating biases, which includes the sub-threshold, staircase operating, and tunnelling breakdown regimes. To better understand our noise model, a detailed discussion of our proposed noise model for an $n$-layer graded-bandgap APD in its staircase operating regime is presented and compared with the two previous noise models. From the discussion, the expressions for the staircase gain of the multistep staircase APD obtained using all three models are equivalent. Moreover, comparing the previous noise models, both models provided equivalent noise expressions. However, comparing our proposed and previous noise models \cite{ref35,ref39,ref41,ref43,ref45}, values of the noise power ratio and excess noise factors of the $n$-step staircase APDs obtained using our model are comparatively greater when $n>1$ and $p\neq 0, 1$. Nevertheless, our noise model and the expression for total excess noise factors of the multistep staircase APDs agree with Bangera's total noise factor of $n$-stage cascade networks \cite{ref48}. Since, Bangera's total noise factor expression for $n$-stage cascade networks \cite{ref48} is a correction to Friis's formula \cite{ref46,ref47}, especially for multistage networks with $n\ge2$; we conclude that our new noise model for $n$-step staircase APDs is an improved noise model.

\section*{Acknowledgments}
A.E.B. would like to thank the Indian Institute of Technology Bombay for their support.

\section*{Appendices}
\appendix

\section{Derivation of the expression for the noise power ratio of the newly proposed noise model and its proof}
\label{Appx_1}

Consider that an electron at the input of step `$x$' would generate only one extra electron when ionized and no extra electrons if not ionized. Then, $X_x$ is a random variable for multiplication at step `$x$' with just two possibilities 0 or 1 with probabilities $(1-p_x)$ or $p_x$, respectively.

\begin{equation}
\label{Appendix_eqn_1}
\begin{aligned}
X_x \sim 
\begin{cases}
0 & (1-p_{x}) \\
& [i.e., \text{No Ionization at step `}x\text{'} \implies X_x=0]\\
1 & p_{x} \\ 
& [i.e., \text{Ionization at step `}x\text{'} \ni \text{1 e$^-$ generated}]\\
\end{cases}
\nonumber
\end{aligned}
\end{equation}

\subsection{Solution for a 1-step staircase APD}

\begin{equation}
\label{Appendix_eqn_3}
\begin{aligned}
(1+X_1)^2 \sim 
\begin{cases}
1  & (1-p_{1}) \\
4  & p_{1} \\ 
\end{cases}
\nonumber
\end{aligned}
\end{equation}

Then, the $\text{NPR}_{\text{S}_n}^\text{new}$ of only a single-step staircase APD is,

\begin{equation}
\label{Appendix_eqn_4}
\begin{aligned}
\text{NPR}_{\text{S}_1}^\text{new} &= \Big\langle (1+X_1)^2 \Big\rangle \\
&= (1-p_1) + 4p_1 \\
&= (1-p_1) \Bigg\{ 1+ (2^2) \biggl[ \frac{p_1}{(1-p_1)} \biggr] \Bigg\} \\
&= 1+3p_1
\nonumber
\end{aligned}
\end{equation}

If the ionization probability $p_1=p$, then $\text{NPR}_{\text{S}_n}^\text{new}$ of the staircase APD with number of steps $n=1$ may be rewritten as,

\begin{equation}
\label{Appendix_eqn_5}
\begin{aligned}
\text{NPR}_{\text{S}_1}^\text{new} = 1+^1C_1(3p) = \sum_{i=0}^{1}~^1C_i(3p)^i
\nonumber
\end{aligned}
\end{equation}

\subsection{Solution for a 2-step staircase APD}

\begin{equation}
\label{Appendix_eqn_6}
\begin{aligned}
\Bigl((1+X_1)^2(1+X_2)^2 \Bigr) \sim 
\begin{cases}
1 & (1-p_1)(1-p_2) \\
4 & p_1(1-p_2) \\ 
4 & (1-p_1)p_2 \\
16 & p_1p_2 \\ 
\end{cases}
\nonumber
\end{aligned}
\end{equation}

Then, the $\text{NPR}_{\text{S}_n}^\text{new}$ of a 2-step staircase APD is,

\begin{equation}
\label{Appendix_eqn_7}
\begin{aligned}
\text{NPR}_{\text{S}_2}^\text{new} &= \Big\langle (1+X_1)^2(1+X_2)^2 \Big\rangle \\
&= (1-p_1)(1-p_2) + 4p_1(1-p_2) + 4(1-p_1)p_2 \\
&~+ 16p_1p_2 \\
&= (1-p_1)(1-p_2) \Biggl\{ 1+ (2^2) \biggl[ \frac{p_1}{(1-p_1)} + \frac{p_2}{(1-p_2)} \biggr] \\
&~+ (2^2)^2 \biggl[ \frac{p_1}{(1-p_1)} \frac{p_2}{(1-p_2)} \biggr] \Biggr\} \\
&= 1+3 \{p_1 + p_2 \} + 9 \{ p_1p_2 \}
\nonumber
\end{aligned}
\end{equation}

If ionization probabilities at both the steps are equal \textit{i.e.}, $p_1=p_2=p$, then $\text{NPR}_{\text{S}_n}^\text{new}$ of the staircase APD with number of steps $n=2$ may be rewritten as,

\begin{equation}
\label{Appendix_eqn_8}
\begin{aligned}
\text{NPR}_{\text{S}_2}^\text{new} = 1+^2C_1 (3p) +^2C_2 (9p^2) = \sum_{i=0}^{2} ~^2C_i(3p)^i
\nonumber
\end{aligned}
\end{equation}

\subsection{Solution for a 3-step staircase APD}

\begin{equation}
\label{Appendix_eqn_9}
\begin{aligned}
\begin{aligned}
\Bigl((1+X_1)^2 (1+X_2)^2 \\
(1+X_3)^2 \Bigr) 
\end{aligned}
\sim 
\begin{cases}
1 & (1-p_1)(1-p_2)(1-p_3) \\
4 & p_1(1-p_2)(1-p_3) \\ 
4 & (1-p_1)p_2(1-p_3) \\
4 & (1-p_1)(1-p_2)p_3 \\
16 & p_1p_2(1-p_3) \\
16 & p_1(1-p_2)p_3 \\ 
16 & (1-p_1)p_2p_3 \\
64 & p_1p_2p_3 \\
\end{cases}
\nonumber
\end{aligned}
\end{equation}

Then, the $\text{NPR}_{\text{S}_n}^\text{new}$ of a 3-step staircase APD is,

\begin{equation}
\label{Appendix_eqn_10}
\begin{aligned}
\text{NPR}_{\text{S}_3}^\text{new} &= \Big\langle (1+X_1)^2(1+X_2)^2(1+X_3)^2 \Big\rangle \\
&= (1-p_1)(1-p_2)(1-p_3) + 4p_1(1-p_2)(1-p_3) \\
&~+ 4(1-p_1)p_2(1-p_3) + 4(1-p_1)(1-p_2)p_3 \\
&~+ 16p_1p_2(1-p_3) + 16p_1(1-p_2)p_3 \\
&~+ 16(1-p_1)p_2p_3 + 64p_1p_2p_3 \\
&= (1-p_1)(1-p_2)(1-p_3) \Biggl\{ 1+ (2^2) \biggl[ \frac{p_1}{(1-p_1)} \\
&~+ \frac{p_2}{(1-p_2)} + \frac{p_3}{(1-p_3)} \biggr] + (2^2)^2 \biggl[ \frac{p_1}{(1-p_1)} \frac{p_2}{(1-p_2)} \\
&~+ \frac{p_1}{(1-p_1)} \frac{p_3}{(1-p_3)} + \frac{p_2}{(1-p_2)} \frac{p_3}{(1-p_3)} \biggr] \\
&~+ (2^2)^3 \biggl[ \frac{p_1}{(1-p_1)} \frac{p_2}{(1-p_2)} \frac{p_3}{(1-p_3)} \biggr] \Biggr\} \\
&= 1+3 \{p_1 + p_2 + p_3 \} + 9 \{ p_1p_2 + p_1p_3 + p_2p_3\} \\
&~+ 27 \{ p_1p_2p_3 \}
\nonumber
\end{aligned}
\end{equation}

If ionization probabilities at all the steps are equal \textit{i.e.}, $p_1=p_2=p_3=p$, then $\text{NPR}_{\text{S}_n}^\text{new}$ of the staircase APD with number of steps $n=3$ may be rewritten as,

\begin{equation}
\label{Appendix_eqn_11}
\begin{aligned}
\text{NPR}_{\text{S}_3}^\text{new} &= 1+~^3C_1 (3p) +~^3C_2 (9p^2) +~^3C_3 (27p^3) \\
&= \sum_{i=0}^{3} ~^3C_i(3p)^i
\nonumber
\end{aligned}
\end{equation}

\subsection{Solution for an \textit{n}-step staircase APD}

\begin{equation}
\label{Appendix_eqn_12}
\begin{aligned}
\prod_{x=1}^{n}(1+X_x)^2 \sim 
\begin{cases}
1 & \prod_{x=1}^{n}(1-p_x) \\
4 & p_1\prod_{x\forall n;x\neq 1}(1-p_x) \\ 
4 & p_2\prod_{x\forall n;x\neq 2}(1-p_x) \\
\vdots & \vdots \\
4 & p_n\prod_{x\forall n;x\neq n}(1-p_x) \\
16 & p_1p_2\prod_{x\forall n;x\neq 1,2}(1-p_x) \\
16 & p_1p_3\prod_{x\forall n;x\neq 1,3}(1-p_x) \\ 
\vdots & \vdots \\
(2^2)^n & \prod_{x=1}^{n}p_x \\
\end{cases}
\nonumber
\end{aligned}
\end{equation}

Then, the $\text{NPR}_{\text{S}_n}^\text{new}$ of an $n$-step staircase APD is,

\begin{equation}
\label{Appendix_eqn_13_1}
\begin{aligned}
\text{NPR}_{\text{S}_n}^\text{new} &= \bigg\langle \prod_{x=1}^{n}(1+X_x)^2 \bigg\rangle \\
&= \prod_{x=1}^{n}(1+p_x) \Biggl\{ 1+2^2 \Biggl[ \sum_{j_1=1}^{n}\frac{p_{j_1}}{(1-p_{j_1})} \Biggr]\\
&\hspace{0.2cm}+(2^2)^2 \Biggl[ \sum_{j_1=1}^{n-1}\frac{p_{j_1}}{(1-p_{j_1})} \biggl[ \sum_{j_2=j_1+1}^{n}\frac{p_{j_2}}{(1-p_{j_2})} \biggr] \Biggr]\\
&\hspace{0.2cm}+(2^2)^3 \Biggl[ \sum_{j_1=1}^{n-2}\frac{p_{j_1}}{(1-p_{j_1})} \biggl[ \sum_{j_2=j_1+1}^{n-1}\frac{p_{j_2}}{(1-p_{j_2})}\\
&\hspace{0.3cm} \Bigl[ \sum_{j_3=j_2+1}^{n}\frac{p_{j_3}}{(1-p_{j_3})} \Bigr] \biggr] \Biggr] + ... +(2^2)^n \Biggl[ \prod_{j_n=1}^{n}p_{j_n}\Biggr]\Biggr\} 
\nonumber
\end{aligned}
\end{equation} 

\begin{equation}
\label{Appendix_eqn_13_2}
\begin{aligned}
\hspace{0.6cm}&=\prod_{x=1}^{n}(1+p_x) \Biggl\{ 1+ \sum_{i=1}^{n}(2^2)^i \Biggl[ \sum_{j_1=1}^{[n-(i-1)]}\sum_{j_2=j_1+1}^{[n-(i-2)]} ...\\
&\hspace{1cm} \sum_{j_i=j_{i-1}+1}^{n}\frac{p_{j_1}}{(1-p_{j_1})}\frac{p_{j_2}}{(1-p_{j_2})} ... \frac{p_{j_i}}{(1-p_{j_i})} \Biggr]\Biggr\}
\nonumber
\end{aligned}
\end{equation} 

 Solving the above expression, $\text{NPR}_{\text{S}_n}^\text{new}$ may also be formulated as, 

\begin{equation}
\label{Appendix_eqn_14}
\begin{aligned}
\text{NPR}_{\text{S}_n}^\text{new} &= 1+ \sum_{i=1}^{n}(3)^i \Biggl[ \sum_{j_1=1}^{[n-(i-1)]}\sum_{j_2=j_1+1}^{[n-(i-2)]} ...\\
&\hspace{1cm} \sum_{j_i=j_{i-1}+1}^{n}p_{j_1}p_{j_2} ... p_{j_i}\Biggr]
\nonumber
\end{aligned}
\end{equation}

If ionization probabilities at all the steps are equal \textit{i.e.}, $p_1=p_2=p_3=...=p_n=p$, then $\text{NPR}_{\text{S}_n}^\text{new}$ of the staircase APD with number of steps `$n$' may be rewritten as,

\begin{equation}
\label{Appendix_eqn_15}
\begin{aligned}
\text{NPR}_{\text{S}_n}^\text{new} = 1+\sum_{i=1}^{n}~^nC_i(3p)^i=\sum_{i=0}^{n}~^nC_i(3p)^i
\nonumber 
\end{aligned}
\end{equation}

\section{Illustrations: Previous noise models follow Friis's total noise factor formula for \textit{n}-stage cascade networks}
\label{Appx_2}

Friis's total noise factor formula for cascade networks \cite{ref46,ref47} is given by,

\begin{equation}
\label{Appendix_B_eqn_1}
F_{\text{T}_n}^{\text{Friis}} = F_1^{\text{Friis}}+\sum_{x=2}^{n}\left(\frac{F_x^{\text{Friis}}-1}{\prod_{y=1}^{(x-1)}M_y}\right)
\nonumber
\end{equation}

Moreover, the Friis noise factor at the $x$-th stage of a cascade network is,

\begin{equation}
\label{Appendix_B_eqn_2}
F_x^{\text{Friis}} = 1+\frac{N_{a(x)}}{N_iM_x} 
\nonumber
\end{equation}

Where $N_{a(x)}$ is an externally added noise at the output of the $x$-th stage or $x$-th step. However, the above expression for $F_x^{\text{Friis}}$ is valid only when all the multiplication probabilities are unity (or the multiplication gain is deterministic). Here, if $N_{a(x)}=0$, then $F_x^{\text{Friis}} = 1$ for all stages or steps. However, if the multiplication probabilities are less than unity and non-zero, then $F_x^{\text{Friis}}$ will be greater than $1$. \\

In the case of solid state devices such as staircase APDs, since there is no external noise added at every step (\textit{i.e.}, $N_{a(x)}=0$) and if all the stepwise ionization or multiplication probabilities of an $n$-step staircase APD are equal  \textit{i.e.}, $p_1=p_2=...=p_n=p$; we could consider that all its stepwise excess noise factors will also be equal. \\

According to previous models \cite{ref35,ref39,ref41,ref43,ref45}, if `$p$' is the ionization probability at each step, then the excess noise factor $F(M_\text{S})^\text{prev}$ of an $n$-step staircase APD is given by, 

\begin{equation}
\label{Appendix_B_eqn_3}
\begin{aligned}
F(M_\text{S})^\text{prev}&=1+\frac{(1-p)(1-(1+p)^{-n})}{(1+p)} = F_{\text{T}_n}^\text{prev}
\nonumber
\end{aligned}
\end{equation} 

The total gain of the $n$-step staircase APD in terms of `$p$' will be $M_\text{S}^\text{prev}=(1+p)^n$. \\

\textbf{Illustration 1:} If $p=0.3$ and $n=1$, then substituting the values of `$p$' and `$n$,' we get the total excess noise factor of a 1-step staircase APD as $F_{\text{T}_1}^\text{prev}=1.12426$. Since this corresponds to a staircase APD with only one step, it could be considered that the stepwise excess noise factor at step one of a 1-step staircase APD is $F_1=F_{\text{T}_1}^\text{prev}=1.12426=F_{\text{T}_1}^{\text{Friis}}$. \\

\textbf{Illustration 2:} Similarly, if $p=0.3$ and $n=2$, the total excess noise factor of a 2-step staircase APD will be $F_{\text{T}_2}^\text{prev}=1.12426$. Since we have considered the same values of `$p$' as in the previous illustration, we get $F_2=F_1=1.12426$. Therefore, substituting $F_2=F_1=1.12426$ in Friis's equation, we get $F_{\text{T}_2}^{\text{Friis}} = F_1 + \frac{(F_2-1)}{M_1} = F_1 + \frac{(F_2-1)}{(1+p)} = 1.12426$. \\

\textbf{Illustration 3:} Similarly, if $p=0.3$ and $n=3$, the total excess noise factor of a 3-step staircase APD will be $F_{\text{T}_3}^\text{prev}=1.293372$.  Since we have considered the same values of `$p$' as in the previous illustration, we get $F_3=F_2=F_1=1.12426$. Therefore, substituting $F_3=F_2=F_1=1.12426$ in Friis's equation, we get $F_{\text{T}_3}^{\text{Friis}} = F_1 + \frac{(F_2-1)}{M_1} + \frac{(F_3-1)}{M_1M_2} = F_1 + \frac{(F_2-1)}{(1+p)} + \frac{(F_3-1)}{(1+p)(1+p)} = 1.293372$. \\

Thus, we conclude that the previous noise models are in accordance with Friis's total noise factor formula for cascade networks.

\section{Illustrations: Our noise model follows Bangera's total noise factor expression for \textit{n}-stage cascade networks}
\label{Appx_3}

Bangera's total noise factor of $n$-stage cascade networks \cite{ref48} given by, 

\begin{equation}
\label{Appendix_C_eqn_1}
F_{\text{T}_n}^{\text{Cor}} = \prod_{x=1}^{n}F_x^{\text{Cor}} 
\nonumber
\end{equation}

Moreover, Bangera's formula for the stage-wise noise factor at the $x$-th stage or $x$-th step ($F_x^{\text{Cor}}$) is,

\begin{equation}
\label{Appendix_C_eqn_2}
\begin{aligned}
F_x^{\text{Cor}} = 1+\frac{N_{a(x)}}{N_i\prod_{j=1}^{x}M_j+\sum_{k=1}^{(x-1)}\{N_{a(k)}\prod_{l=k+1}^{x}M_l\}}
\nonumber
\end{aligned}
\end{equation}

Where $N_{a(x)}$ is an externally added noise at the output of the $x$-th stage or $x$-th step. Similar to Friis conditions, the above expression for $F_x^{\text{Cor}}$ is valid only when all the multiplication probabilities are unity (or the multiplication gain is deterministic). Here, if $N_{a(x)}=0$, then $F_x^{\text{Cor}} = 1$ for all stages or steps. However, if the multiplication probabilities are less than unity and non-zero, then $F_x^{\text{Cor}}$ will be greater than $1$. \\

Again, in the case of solid state devices such as staircase APDs, since there is no external noise added at every step (\textit{i.e.}, $N_{a(x)}=0$) and if all the stepwise ionization or multiplication probabilities of an $n$-step staircase APD are equal \textit{i.e.}, $p_1=p_2=...=p_n=p$; we could consider that all its stepwise excess noise factors will also be equal. \\

According to our new model, if `$p$' is the ionization probability at each step, then the excess noise factor $F(M_\text{S})^\text{new}$ of an $n$-step staircase APD is given by, 

\begin{equation}
\label{Appendix_C_eqn_3}
F(M_\text{S})^\text{new} = \frac{\sum_{i=0}^{n}~^nC_i(3p)^i}{(1+p)^{2n}} = F_{\text{T}_n}^\text{new}
\nonumber
\end{equation}

The total gain of the $n$-step staircase APD in terms of `$p$' will be $M_\text{S}^\text{new}=(1+p)^n$. \\

\textbf{Illustration 1:} If $p=0.3$ and $n=1$, then substituting the values of `$p$' and `$n$' we get the total excess noise factor of a 1-step staircase APD as $F_{\text{T}_1}^\text{new}=1.12426$. Since this corresponds to a staircase APD with only one step, it could be considered that the stepwise excess noise factor at step one of a 1-step staircase APD is $F_1=F_{\text{T}_1}^\text{new}=1.12426=F_{\text{T}_1}^{\text{Cor}}$. This value matches the total excess noise factor of a 1-step staircase APD obtained using the previous models. \\

\textbf{Illustration 2:} Similarly, if $p=0.3$ and $n=2$, the total excess noise factor of a 2-step staircase APD will be $F_{\text{T}_2}^\text{new}=1.26396$. Since we have considered the same values of `$p$' as in the previous illustration, we get $F_2=F_1=1.12426$. Therefore, substituting $F_2=F_1=1.12426$ in Bangera's equation, we get $F_{\text{T}_2}^{\text{Cor}} = F_1F_2 = 1.26396$. \\

\textbf{Illustration 3:} Similarly, if $p=0.3$ and $n=3$, the total excess noise factor of a 3-step staircase APD will be $F_{\text{T}_3}^\text{new}=1.42102$.  Since we have considered the same values of `$p$' as in the previous illustration, we get $F_3=F_2=F_1=1.12426$. Therefore, substituting $F_3=F_2=F_1=1.12426$ in Bangera's equation, we get $F_{\text{T}_3}^{\text{Cor}} = F_1F_2F_3 = 1.42102$. \\
 
From the above illustrations, when $n\geq2$, the total excess noise factors of $n$-step staircase APDs obtained using our model are greater than the corresponding total excess noise factors obtained using the previous models. Moreover, we conclude that our noise model follows Bangera's total noise factor expression for cascade networks.

\bibliographystyle{unsrt}
\bibliography{BibRef}

\begin{thebibliography}{10}

\bibitem{ref1}
Hamamatsu Photonics.
\newblock {\em Photomultiplier Tubes: {P}hotomultiplier Tubes and Related
  Products}, 2016.

\bibitem{ref2}
R.~W. Engstrom, F.~A. Helvy, H.~R. Krall, T.~T. Lewis, W.~D. Lindley, R.~U.
  Martinelli, R.~M. Matheson, A.~G. Nekut, D.~E. Persyk, R.~M. Shaffer, and
  A.~H. Sommer.
\newblock {\em Photomultiplier Manual}.
\newblock RCA Corporation, 1970.

\bibitem{ref3}
George~Ashmun Morton.
\newblock Photomultipliers for scintillation counting.
\newblock {\em RCA Rev.}, 10:525--553, 1949.

\bibitem{ref4}
Richard K.~P. Benninger, William~J. Ashby, Elisabeth~A. Ring, and David~W.
  Piston.
\newblock A single-photon-counting detector for increased sensitivity in
  two-photon laser scanning microscopy.
\newblock {\em Opt. Lett.}, 33(24):2895--2897, 2008.

\bibitem{ref5}
PerkinElmer, Inc.
\newblock {\em {LAMBDA 650/750/850/950/1050} Accessories for {LAMBDA} Series},
  2012.

\bibitem{ref6}
Frank Padera.
\newblock {\em {Characterization of Custom Light Sources Using the {LAMBDA
  650/750/850/950/1050 UV/Vis and UV/Vis/NIR} Research Spectrometers}}.
\newblock PerkinElmer, Inc. Shelton, CT, 2016.

\bibitem{ref7}
{The ATLAS Collaboration}.
\newblock The {ATLAS} experiment at the {CERN} large hadron collider.
\newblock {\em J. Instrum.}, 3(08):S08003--S08003, 2008.

\bibitem{ref8}
Yoichiro Suzuki.
\newblock The {Super-Kamiokande} experiment.
\newblock {\em The European Physical Journal C}, 79(4):1434--6052, 2019.

\bibitem{ref9}
C.~W. Oatley.
\newblock The scanning electron microscope.
\newblock {\em Sci. Prog. (1933- )}, 54(216):483--495, 1966.

\bibitem{ref10}
Karsten K{\"o}nig.
\newblock Clinical multiphoton tomography.
\newblock {\em J. Biophotonics}, 1(1):13--23, 2008.

\bibitem{ref11}
H.~Kume, S.~Suzuki, and K.~Oba.
\newblock Recent development of photomultiplier tubes for nuclear and medical
  applications.
\newblock {\em IEEE Trans. Nucl. Sci.}, 32(1):355--359, 1985.

\bibitem{ref12}
Silvano Donati and Tiziana Tambosso.
\newblock Single-photon detectors: From traditional {PMT} to solid-state
  {SPAD}-based technology.
\newblock {\em IEEE J. Sel. Top. Quantum Electron.}, 20(6):204--211, 2014.

\bibitem{ref13}
Robert~H. Hadfield.
\newblock Single-photon detectors for optical quantum information applications.
\newblock {\em Nat. Photonics}, 3(12):696--705, 2009.

\bibitem{ref14}
Z.~L. Yuan, B.~E. Kardynal, A.~W. Sharpe, and A.~J. Shields.
\newblock High speed single photon detection in the near infrared.
\newblock {\em Appl. Phys. Lett.}, 91(4):041114, 2007.

\bibitem{ref15}
Xudong Jiang, Mark~A. Itzler, Rafael Ben-Michael, and Krystyna Slomkowski.
\newblock {InGaAsP}-{InP} avalanche photodiodes for single photon detection.
\newblock {\em IEEE J. Sel. Top. Quantum Electron.}, 13(4):895--905, 2007.

\bibitem{ref16}
Jun Zhang, Rob Thew, Claudio Barreiro, and Hugo Zbinden.
\newblock Practical fast gate rate {InGaAs}/{InP} single-photon avalanche
  photodiodes.
\newblock {\em Appl. Phys. Lett.}, 95(9):091103, 2009.

\bibitem{ref17}
B.~E. Kardyna{\l}, Z.~L. Yuan, and A.~J. Shields.
\newblock An avalanche-photodiode-based photon-number-resolving detector.
\newblock {\em Nat. Photonics}, 2(7):425--428, 2008.

\bibitem{ref18}
D.~Dunai, S.~Zoletnik, J.~S{\'a}rk{\"o}zi, and A.~R. Field.
\newblock Avalanche photodiode based detector for beam emission spectroscopy.
\newblock {\em Rev. Sci. Instrum}, 81(10):103503, 2010.

\bibitem{ref19}
M~D~C Whitaker, G~Lioliou, A~B Krysa, and A~M Barnett.
\newblock {Al}$_{0.6}${Ga}$_{0.4}${As} {X}-ray avalanche photodiodes for
  spectroscopy.
\newblock {\em Semicond Sci Technol.}, 35(9):095026, 2020.

\bibitem{ref20}
A.~Auckloo, J.S. Cheong, X.~Meng, C.H. Tan, J.S. Ng, A.~Krysa, R.C. Tozer, and
  J.P.R. David.
\newblock {Al}$_{0.52}${In}$_{0.48}${P} avalanche photodiodes for soft {X}-ray
  spectroscopy.
\newblock {\em J. Instrum.}, 11(03):P03021--P03021, 2016.

\bibitem{ref21}
Claudio Bruschini, Harald Homulle, Ivan~Michel Antolovic, Samuel Burri, and
  Edoardo Charbon.
\newblock Single-photon avalanche diode imagers in biophotonics: review and
  outlook.
\newblock {\em Light Sci. Appl.}, 8(1):87, 2019.

\bibitem{ref22}
Cesar Bartolo-Perez, Soroush Chandiparsi, Ahmed~S. Mayet, Hilal Cansizoglu,
  Yang Gao, Wayesh Qarony, Ahasan AhAmed, Shih-Yuan Wang, Simon~R. Cherry,
  M.~Saif Islam, and Gerard Ari{\~n}o-Estrada.
\newblock Avalanche photodetectors with photon trapping structures for
  biomedical imaging applications.
\newblock {\em Opt. Express}, 29(12):19024--19033, 2021.

\bibitem{ref23}
R.J. McIntyre.
\newblock Multiplication noise in uniform avalanche diodes.
\newblock {\em IEEE Trans. Electron Devices}, ED-13(1):164--168, 1966.

\bibitem{ref24}
M.~Teich, K.~Matsuo, and B.~Saleh.
\newblock Excess noise factors for conventional and superlattice avalanche
  photodiodes and photomultiplier tubes.
\newblock {\em IEEE J. Quantum Electron}, 22(8):1184--1193, 1986.

\bibitem{ref25}
H.A. Haus.
\newblock {\em Electromagnetic Noise and Quantum Optical Measurements}.
\newblock Advanced Texts in Physics. Springer-Verlag Berlin Heidelberg,
  Heidelberg, Germany, 2000.

\bibitem{ref26}
Andrew~S. Huntington.
\newblock {\em {InGaAs} Avalanche Photodiodes for Ranging and Lidar}.
\newblock Woodhead Publishing Series in Electronic and Optical Materials.
  Woodhead Publishing is an imprint of Elsevier, Duxford, United Kingdom, 2020.

\bibitem{ref27}
J.~C. Blakesley, P.~See, A.~J. Shields, B.~E. Kardyna\l{}, P.~Atkinson,
  I.~Farrer, and D.~A. Ritchie.
\newblock Efficient single photon detection by quantum dot resonant tunneling
  diodes.
\newblock {\em Phys. Rev. Lett.}, 94:067401, 2005.

\bibitem{ref28}
Solomon Assefa, Fengnian Xia, and Yurii~A. Vlasov.
\newblock Reinventing germanium avalanche photodetector for nanophotonic
  on-chip optical interconnects.
\newblock {\em Nature}, 464(7285):80--84, 2010.

\bibitem{ref29}
Alan~C. Farrell, Xiao Meng, Dingkun Ren, Hyunseok Kim, Pradeep Senanayake,
  Nick~Y. Hsieh, Zixuan Rong, Ting-Yuan Chang, Khalifa~M. Azizur-Rahman, and
  Diana~L. Huffaker.
\newblock {InGaAs}-{GaAs} nanowire avalanche photodiodes toward single-photon
  detection in free-running mode.
\newblock {\em Nano Lett.}, 19(1):582--590, 2019.

\bibitem{ref30}
Oliver Hayden, Ritesh Agarwal, and Charles~M. Lieber.
\newblock Nanoscale avalanche photodiodes for highly sensitive and spatially
  resolved photon detection.
\newblock {\em Nat. Mater.}, 5(5):352--356, 2006.

\bibitem{ref31}
Min Ren, Scott~J. Maddox, Madison~E. Woodson, Yaojia Chen, Seth~R. Bank, and
  Joe~C. Campbell.
\newblock {AlInAsSb} separate absorption, charge, and multiplication avalanche
  photodiodes.
\newblock {\em Appl. Phys. Lett.}, 108(19):191108, 2016.

\bibitem{ref32}
Xin Yi, Shiyu Xie, Baolai Liang, Leh~W. Lim, Jeng~S. Cheong, Mukul~C. Debnath,
  Diana~L. Huffaker, Chee~H. Tan, and John P.~R. David.
\newblock Extremely low excess noise and high sensitivity
  {AlAs}$_{0.56}${Sb}$_{0.44}$ avalanche photodiodes.
\newblock {\em Nat. Photonics}, 13(10):683--686, 2019.

\bibitem{ref33}
Andrew~H. Jones, Stephen~D. March, Seth~R. Bank, and Joe~C. Campbell.
\newblock Low-noise high-temperature {AlInAsSb}/{GaSb} avalanche photodiodes
  for 2-\textmu{m} applications.
\newblock {\em Nat. Photonics}, 14(9):559--563, 2020.

\bibitem{ref34}
R.~Chin, N.~Holonyak, G.E. Stillman, J.Y. Tang, and K.~Hess.
\newblock Impact ionisation in multilayered heterojunction structures.
\newblock {\em Electron. Lett.}, 16:467--469, 1980.

\bibitem{ref35}
G.F. Williams, F.~Capasso, and W.T. Tsang.
\newblock The graded bandgap multilayer avalanche photodiode: A new low-noise
  detector.
\newblock {\em IEEE Electron Device Lett.}, 3(3):71--73, 1982.

\bibitem{ref36}
F.~Capasso.
\newblock Avalanche photodiodes with enhanced ionization rates ratio: Towards a
  solid state photomultiplier.
\newblock {\em IEEE Trans. Nucl. Sci.}, 30(1):424--428, 1983.

\bibitem{ref37}
G.~Ripamonti, F.~Capasso, A.L. Hutchinson, D.J. Muehlner, J.F. Walker, and R.J.
  Malik.
\newblock Realization of a staircase photodiode: Towards a solid-state
  photomultiplier.
\newblock {\em Nucl. Instrum. Methods Phys. Res. A}, 288(1):99--103, 1990.

\bibitem{ref38}
John David.
\newblock Photodetectors: The staircase photodiode.
\newblock {\em Nat. Photonics}, 10(6):364--366, 2016.

\bibitem{ref39}
Stephen~D. March, Andrew~H. Jones, Joe~C. Campbell, and Seth~R. Bank.
\newblock Multistep staircase avalanche photodiodes with extremely low noise
  and deterministic amplification.
\newblock {\em Nat. Photonics}, 15(6):468--474, 2021.

\bibitem{ref40}
Min Ren, Scott Maddox, Yaojia Chen, Madison Woodson, Joe~C. Campbell, and Seth
  Bank.
\newblock {AlInAsSb}/{GaSb} staircase avalanche photodiode.
\newblock {\em Appl. Phys. Lett.}, 108(8):081101, 2016.

\bibitem{ref41}
F.~Capasso, Won-Tien Tsang, and G.F. Williams.
\newblock Staircase solid-state photomultipliers and avalanche photodiodes with
  enhanced ionization rates ratio.
\newblock {\em IEEE Trans. Electron Devices}, 30(4):381--390, 1983.

\bibitem{ref42}
K.~Matsuo, M.C. Teich, and B.E.A. Saleh.
\newblock Noise properties and time response of the staircase avalanche
  photodiode.
\newblock {\em IEEE Trans. Electron Devices}, 32(12):2615--2623, 1985.

\bibitem{ref43}
A.~Pilotto, P.~Palestri, L.~Selmi, M.~Antonelli, F.~Arfelli, G.~Biasiol,
  G.~Cautero, F.~Driussi, R.~H. Menk, C.~Nichetti, and T.~Steinhartova.
\newblock A new expression for the gain-noise relation of single-carrier
  avalanche photodiodes with arbitrary staircase multiplication regions.
\newblock {\em IEEE Trans. Electron Devices}, 66(4):1810--1814, 2019.

\bibitem{ref44}
PerkinElmer, Inc.
\newblock {\em Avalanche Photodiodes: A User's Guide}, 2003.

\bibitem{ref45}
A.~van~der Ziel, Y.J. Yu, G.~Bosman, and C.M. Van~Vliet.
\newblock Two simple proofs of {C}apasso's excess noise factor {F}$_{N}$of an
  ideal ${N}$-stage staircase multiplier.
\newblock {\em IEEE Trans. Electron Devices}, 33(11):1816--1817, 1986.

\bibitem{ref46}
H.T. Friis.
\newblock Noise figures of radio receivers.
\newblock {\em Proc. IRE}, 32(7):419--422, 1944.

\bibitem{ref47}
H.A. Haus.
\newblock The noise figure of optical amplifiers.
\newblock {\em IEEE Photon. Technol. Lett.}, 10(11):1602--1604, 1998.

\bibitem{ref48}
Ankitha~E. Bangera.
\newblock Correction to {Friis} noise factors.
\newblock {P}reprint at \url{https://doi.org/10.48550/arXiv.2209.00287}, 2022.

\end{thebibliography}

\end{document}